\documentclass[preprint,prd,showpacs,showkeys,floatfix]{revtex4}
\usepackage{latexsym,amssymb,amsmath,amsfonts}

\renewcommand{\ol}[1]{{\overline{#1}}}

\begin{document}

\date{\today}
\title{New symmetry current for massive spin-$\frac{3}{2}$ fields}
\author{Terry~Pilling}
\email{terry@member.ams.org}
\affiliation{Bogoliubov Laboratory of Theoretical Physics,
Joint Institute for Nuclear Research,
141980 Dubna, Russia\\
and\\
Institute of Theoretical and Experimental Physics,
117218 B. Cheremushkinskaya 25, Moscow, Russia}


\begin{abstract}
We present several new results which will be of value to theorists
working with massive spin-$\frac{3}{2}$ vector-spinor fields as found,
for example, in low and intermediate energy hadron physics and also linearized
supergravity. The most general lagrangian and propagator for
a vector-spinor field in $d$-dimensions is given. It is shown that the 
observables of the theory are invariant under a novel continuous symmetry group 
which is also extended to an algebra. 
A new technique is developed for exploring the consequences of the symmetry
and a previously unknown conserved vector current and charge are found. 
The current leads to new interactions involving spin-$\frac{3}{2}$ particles 
and may have important experimental consequences. 
\end{abstract}

\maketitle

\section{General Rarita-Schwinger action}

In many instances, a relativistic field theory description of particles of 
spin $\geq 1$ requires the presence of auxiliary lower-spin degrees 
of freedom. This arises from the fact that a massive field of spin $s$ 
has $2s+1$ degrees of freedom, whereas the description in terms of Lorentz 
covariant tensors and spinors may require more than $2s+1$ components when $s \geq 1$.
Already when $s = 1$ an extra spin-0 field is needed 
and the situation becomes more interesting when 
$s \geq \frac{3}{2}$ since several extra fields are required and there 
appears a new symmetry of the theory. The symmetry is 
realized by a continuous group of transformations which redefine these lower 
spin auxiliary fields while leaving the observable physics unaffected.
The first non-trivial case where this new symmetry group appears
is that of spin-$\frac{3}{2}$ where two spin-$\frac{1}{2}$ auxiliary fields 
are needed. In this paper we examine some of the consequences of this symmetry.
We work in $d$ spacetime dimensions so that our expressions remain 
general and can be more easily facilitate dimensional regularization in 
hadronic applications. 

The most general lagrangian for a massive vector-spinor field 
in $d$ dimensions is given by \cite{pilling2004}
\begin{equation}
\label{action1}
\mathcal{L} = \ol{\psi}_\alpha 
\left( \Gamma^{\alpha \mu \beta} i \partial_\mu - m \Gamma^{\alpha \beta} \right)
\psi_\beta,
\end{equation}
where $\psi^\alpha$ is a complex vector-spinor field with suppressed 
spinor index \footnote{See \cite{wetterich1983} and references therein 
for properties of spinors in $d$ dimensions, the $d$-dimensional 
Lorentz group and the group of $d$-dimensional general coordinate 
transformations.} and 
\begin{equation}
\begin{split}
\Gamma^{\alpha \mu \beta} &= g^{\alpha \beta} \gamma^\mu - A_1 g^{\mu \beta}
\gamma^\alpha - A_2 g^{\mu \alpha} \gamma^\beta + A_3 \gamma^\alpha \gamma^\mu
\gamma^\beta, \\
\Gamma^{\alpha \beta} &= g^{\alpha \beta} - A_4 \gamma^\alpha \gamma^\beta.
\end{split}
\end{equation}
The coefficients are defined in terms of a complex parameter $a$ by
\begin{equation}
\begin{split}
A_1 &= 1 + \frac{(d-2)}{d}a^*, \quad A_2 = 1 + \frac{(d-2)}{d}a, \\
A_3 &= 1 + \frac{(d-2)}{d} \left[\frac{(d-1)}{d} |a|^2 + a^* + a \right], \\
A_4 &= 1 + \frac{(d-1)}{d} \left[|a|^2 + a^* + a\right].
\end{split}
\end{equation}
This can be written in a simpler way by defining the transformation
$\theta^{\mu \nu}(a)$ as follows
\begin{equation}
\label{pointtrans}
\theta^{\mu \nu}(a) = g^{\mu \nu} + \frac{a}{d} \gamma^\mu \gamma^\nu,  
\quad a \neq -1. 
\end{equation}
Then our action (\ref{action1}) becomes
\begin{equation}
\label{action2}
\mathcal{L} = \ol{\psi}^\alpha \theta_{\alpha \mu}(a^*)
\left(\gamma^{\mu \rho \nu} i \partial_\rho + m \gamma^{\mu \nu}
\right) \theta_{\nu \beta}(a) \psi^\beta,
\end{equation}
where the totally antisymmetric combinations of gamma matrices are given by
$\gamma^{\mu \rho \nu} = \frac{1}{2} \left( \gamma^\mu \gamma^\rho \gamma^\nu 
- \gamma^\nu \gamma^\rho \gamma^\mu \right)$ and $\gamma^{\mu \nu} = 
\frac{1}{2} \left( \gamma^\mu \gamma^\nu - \gamma^\nu \gamma^\mu \right)$.
In the massless limit, the action (\ref{action1}) has the invariance
$\delta \psi^\alpha = \theta^{\alpha \beta} ({\tilde{a}}) \; \partial_\beta \epsilon$ 
for arbitrary spinor $\epsilon$, where the inverse transformation 
$\theta^{\alpha \beta}({\tilde{a}})$ is defined below.
The parameter choice $a=0$ corresponds to the expression commonly 
found \cite{van1981} as the massive gravitino action in linearized supergravity
and in that case the invariance reduces to the usual massless gravitino gauge invariance.
In addition, the choices $a = \frac{d}{1-d}$ and $a = \frac{(A + 1)d}{2-d}$
for real $A$ respectively give the original Rarita-Schwinger action \cite{rarita1941} and 
an expression often seen in nuclear resonance applications.
In fact, all of the vector-spinor lagrangians found in the literature 
are equivalent to the general action for various choices of parameter.
This is as it should be since the general form of the action is dictated
by the principles of quantum field theory \cite{aurilia1969}.
Notice that the choice of parameter giving the supergravity action is such 
that the dimension of spacetime, $d$, does not explicitly appear.
This, along with the antisymmetry of $\gamma^{\mu \rho \nu}$ and 
$\gamma^{\mu \nu}$, make manipulations simpler and results more 
transparent which is why we choose to use this parametrization of the
general action.

\section{Point transformation group}

The transformations given by (\ref{pointtrans})
are sometimes called `point' or `contact' transformations
in the literature. When $a \neq -1$ they form a group with
\begin{equation}
\begin{split}
\theta^{\mu \nu}(a) \theta_{\nu \lambda}(b) 
&= \theta^\mu_{\; \lambda}(a + b + ab) \equiv \theta^\mu_{\; \lambda}(a \circ b), \\
\left(\theta^{\mu \nu}\right)^{-1}(a) &= 
\theta^{\mu \nu}(\frac{-a}{1+a}) \equiv \theta^{\mu \nu}(\tilde{a}),
\end{split}
\end{equation}
where we have defined the `circle' operation $a \circ b = a + b + ab$ and 
also the inverse parameter $\tilde{a} = \frac{-a}{1+a}$.
The transformation becomes singular at the parameter value $a = -1$ as 
can be seen by the fact that $k \circ -1 = -1$ for any $k$, so that
\begin{equation}
\label{singular}
\theta^{\mu \nu}(-1) \theta_{\nu \lambda}(k) = \theta^\mu_{\; \lambda}(-1)
\quad \forall \; k. 
\end{equation}
Interestingly, $a = -1$ gives the additive identity element of
an algebra defined by the following addition rule
\begin{equation}
\begin{split}
\theta_{\mu \nu}(a) + \theta_{\mu \nu}(b) 
&= \theta_{\mu \nu}(a + b + 1), \\
\theta_{\mu \nu}(a) - \theta_{\mu \nu}(b) 
&= \theta_{\mu \nu}(a - b - 1). 
\end{split}
\end{equation}
This addition is, of course, defined {\it modulo zero},
where zero is the two-sided ideal generated by $\theta_{\mu \nu}(-1)$. 
It is easily shown that the multiplication is distributive over addition.

We should mention that one can redefine the parameter in various ways to
make the group more convenient. For example, by shifting the singular 
point of the parameter space to $- \infty$ by letting 
$a \rightarrow e^{\alpha} - 1$ the group becomes \cite{pascalutsa1999}
\begin{equation}
\begin{split}
\theta^{\mu \nu}(\alpha) &= g^{\mu \nu} + \frac{e^{\alpha} -
1}{d} \gamma^\mu \gamma^\nu = e^{\frac{\alpha}{d} \gamma^\mu
\gamma^\nu} \\
\theta^{\mu \nu}(\alpha) \theta_{\nu \lambda}(\beta) 
&= \theta^\mu_\lambda(\alpha + \beta), \\
\left(\theta^{\mu \nu}\right)^{-1}(\alpha) &= \theta^{\mu \nu}(-\alpha).
\end{split}
\end{equation}
An even more convenient redefinition is so that the 
singular point is at 0. One has $a \rightarrow \alpha - 1$ and 
the algebra is then defined by
\begin{eqnarray*}
\theta^{\mu \nu}(\alpha) =& g^{\mu \nu} + \frac{\alpha - 1}{d} \gamma^\mu \gamma^\nu, 
& \text{(definition)}  \\
\theta^{\mu \lambda}(\alpha) \theta_{\lambda}^{\; \; \nu}(\beta) 
=& \theta^{\mu \nu}(\alpha \beta), 
& \text{(multiplication)}  \\
\theta^{\mu \nu}(1) =& g^{\mu \nu}, & \text{(mult. id.)}  \\
\left(\theta^{\mu \nu}\right)^{-1}(\alpha) =& \theta^{\mu \nu}(\frac{1}{\alpha}), 
& \text{(mult. inv.)}  \\
\theta^{\mu \nu}(\alpha) + \theta^{\mu \nu}(\beta) =& \theta^{\mu \nu}(\alpha + \beta), 
& \text{(addition)}  \\
\theta^{\mu \nu}(0) =& g^{\mu \nu} - \frac{1}{d} \gamma^\mu \gamma^\nu, 
& \text{(additive id.)}  \\
\theta^{\mu \nu}(\alpha) - \theta^{\mu \nu}(\beta) =& \theta^{\mu \nu}(\alpha - \beta),
& \text{(additive inv.)}
\end{eqnarray*}
where the addition is again defined modulo the additive identity. 
These redefinitions have advantages in terms of the simplicity of 
algebraic operations; nevertheless, for now we will continue to use the parameter as 
defined by (\ref{pointtrans}).

The path integral is invariant under a global point transformation of the fields
since the functional determinant is trivial and factors out of the integral to be cancelled 
out of the generating functional by the identical factor in the denominator.
Hence there are no path integral anomalies and all physical correlation 
functions are independent of the parameter $a$. 
In the interacting theory the same will be true if one chooses the interaction
to satisfy the same general criteria used in deriving the general free
action: hermiticity, linearity in derivatives and non-singular 
behavior \cite{aurilia1969}.
The full interacting theory will then have the same parameter 
dependence as the free theory, so that a shift of parameter in the 
full action is the same as a point transformation of the fields. 
This independence of correlation functions on the value of the parameter 
can also be seen in the following way. Write the general action (\ref{action2}) as 
\begin{equation} 
\mathcal{L} = \ol{\psi}^\alpha \Lambda_{\alpha \beta}(a) \psi^\beta = \ol{\psi}^\alpha \theta_{\alpha \mu}(a^*) \Lambda^{\mu \nu}_{\text{SG}}
\theta_{\nu \beta}(a) \psi^\beta, 
\end{equation}
where $\Lambda^{\mu \nu}_{\text{SG}}$ is defined by comparison with 
(\ref{action2}). The propagator $\Gamma_{\text{SG}}$ (dropping indices) is the 
inverse of $\Lambda_{\text{SG}}$. The general propagator can be found from 
this by applying point transformations to the relation 
$\Lambda_{\text{SG}} \Gamma_{\text{SG}} = 1$, 
\begin{equation}
\theta(a^*) \Lambda_{\text{SG}} \theta(a) \theta^{-1}(a) \Gamma_{\text{SG}}
\theta^{-1}(a^*) = 1.
\end{equation}
The general propagator is then
\begin{equation}
\label{propagator1}
\Gamma(a) = \theta^{-1}(a) \Gamma_{\text{SG}} \theta^{-1}(a^*).
\end{equation}
Thus the propagator line in a Feynman diagram carries the inverse 
of the transformation on each end. Any interaction vertex which obeys our
general conditions will carry a corresponding transformation to cancel it 
and so all observable quantities will be invariant under a shift of the parameter. 
It can be set to any convenient value in the quantum theory and 
observable physics will be unaffected. 

\section{New symmetry current}

We will now examine one of the ways in which the point transformation invariance 
of the correlation functions can be exploited. 
Since the observables of the quantum theory are independent of the 
parameter choice, we would like to explore the consequences of this invariance using 
Noether's method. However, the classical action is not invariant under
the transformation since the symmetry transformation, $\theta(k)$, of the 
field is {\it non-unitary}, $\theta(k^*) \neq \theta^{-1}(k)$.  
Under a global point transformation, $\psi_\mu \rightarrow \theta_{\mu \nu}(k)\psi^\nu$, 
the lagrangian (\ref{action1}) is not invariant, but transforms as 
$\mathcal{L}(a) \rightarrow \mathcal{L}(a \circ k)$. 
In the absence of interactions the equations of motion are invariant since
the free field equations of motion imply that $\gamma \cdot \psi = 0$ and
this makes the transformation (\ref{pointtrans}) trivial.
However, in the interacting theory this is no longer true. In order to explore
the consequences of the symmetry we will therefore {\it impose it} on the classical 
action by using the following technique. We simply demand that: 
\begin{itemize}
\item{\it classical actions will be considered equivalent if they lead to
the same quantum theory}. 
\end{itemize}
Put in another way this says:
\begin{itemize}
\item{\it classical actions will be considered equivalent if they are related 
by a circle-shift of the parameter}.
\end{itemize}
This makes the point transformations a symmetry of {\it equivalence 
classes} of classical actions and we can use Noether's method to examine 
the consequences. 

For simplicity, we will re-write our action so that it is symmetric 
in derivatives and we will restrict the parameter to be real. 
Thus 
\begin{equation}
\label{symaction}
\mathcal{L}(a) =  \ol{\psi}_\alpha \left[\frac{1}{2} 
\Gamma^{\alpha \rho \beta}(a)i \overset{\leftrightarrow}{\partial}_\rho 
+ m \Gamma^{\alpha \beta}(a) \right] \psi_\beta,
\end{equation}
where we have written 
\begin{equation}
\begin{split}
\Gamma^{\alpha \rho \beta}(a) &=  \theta^{\alpha}_{\; \mu}(a) \gamma^{\mu \rho \nu}
\theta_{\nu}^{\; \beta}(a), \\
\Gamma^{\alpha \beta}(a) &= \theta^{\alpha}_{\; \mu}(a) \gamma^{\mu \nu}
\theta_{\nu}^{\; \beta}(a), \\
\overset{\leftrightarrow}{\partial}_\rho &= 
\overset{\rightarrow}{\partial}_\rho - \overset{\leftarrow}{\partial}_\rho. 
\end{split}
\end{equation}
Under an infinitesimal local point transformation, $\theta(k(x))$, the 
lagrangian varies as
$\mathcal{L}(a) \rightarrow \mathcal{L}(a \circ k) + \delta \mathcal{L}$
where $\delta \mathcal{L}$ contains the derivative acting on the parameter
and $\mathcal{L}(a \circ k)$ is defined in exactly the same way as 
$\mathcal{L}(a)$ in (\ref{symaction}) with the derivatives acting only on the fields 
and not on the parameter. Explicit computation gives
\begin{equation}
\begin{split}
\delta \mathcal{L} &= \frac{i}{2d} \ol{\psi}_\alpha \biggl[ \Gamma^{\alpha \rho
\beta} \gamma_\beta \gamma^\nu - \gamma^\alpha \gamma_\beta \Gamma^{\beta
\rho \nu} \biggr] \psi_\nu \left(\partial_\rho k\right).
\end{split}
\end{equation}
Integrating by parts and discarding the surface term we have
\begin{equation}
\label{deltaL}
\mathcal{L}(a) \rightarrow \mathcal{L}(a \circ k) 
- \frac{1}{2d} \left(\partial_\rho J^\rho\right) k(x),
\end{equation}
where $J^\rho$ is given by
\begin{equation}
\label{current}
J^\rho = i \ol{\psi}_\alpha \left[ \Gamma^{\alpha \rho \beta}(a) \gamma_\beta
\gamma^\nu - \gamma^\alpha \gamma_\beta \Gamma^{\beta \rho \nu}(a) \right] \psi_\nu.
\end{equation}
Our symmetry says that $\mathcal{L}(a) = \mathcal{L}(a \circ k)$ in the limit 
that $k(x)$ becomes constant.
This demands that $\delta \mathcal{L} =0$ in the limit of constant $k(x)$. 
Hence, from (\ref{deltaL}), we find a conserved current, $J^\rho$, associated 
to the global symmetry: $\partial_\rho J^\rho = 0$. 
We see from (\ref{current}) that the current
changes under point transformations by a circle-shift of the parameter
and is therefore invariant according to our symmetry.
We can expand the $\Gamma^{\alpha \rho \beta}$ to find a simpler
expression of the current as follows
\begin{equation}
\begin{split}
J^\rho &= i (1+a) \ol{\psi}_\alpha \left[ \gamma^\alpha g^{\rho \beta} 
- g^{\alpha \rho} \gamma^\beta \right] \psi_\beta, \\
&= i (1+a) \left[\ol{\psi} \cdot \gamma \psi^\rho
- \ol{\psi}^\rho \gamma \cdot  \psi \right].
\end{split}
\end{equation}
Under a transformation, $\theta(k)$, the only change is the coefficient 
$(1+a) \rightarrow (1 + a \circ k)$. The conserved charge is given by
\begin{equation}
\label{charge}
Q = i (1+a) \int d^{d-1} x 
\left[ 
\left(\ol{\psi} \cdot \gamma \right) \psi^0 - \ol{\psi}^0 \left(\gamma
\cdot \psi \right) \right],
\end{equation}
which can be put in a more suggestive form by defining the spin-$\frac{1}{2}$
fields $\chi_1 = \gamma \cdot \psi$ and $\chi_2 = \gamma^0 \psi^0$, so the charge
becomes
\begin{equation}
\label{newEMcharge}
Q = i (1+a) \int d^{d-1} x 
\left[\chi_1^\dagger \chi_2 - \chi_2^\dagger \chi_1 \right].
\end{equation}
Since we have a new conserved current, $J^\rho$, we can couple a 
vector field such as the photon to it as follows
\begin{equation}
\label{newEM}
\begin{split}
\mathcal{L}_{\text{int}} &= g J_\mu A^\mu \\
&= i g (1+a) \left[\ol{\psi} \cdot \gamma \psi_\mu
- \ol{\psi}_\mu \gamma \cdot  \psi \right] A^\mu,
\end{split}
\end{equation}
where $g$ is a coupling constant. 
If this coupling is physically reasonable, then it should, among other
things, have a measurable effect on the magnetic moment of the 
spin-$\frac{3}{2}$ particle.
Furthermore, we also have the usual conserved vector current coming
from electromagnetic gauge symmetry. This is given by
\begin{equation}
j^\mu = \ol{\psi}_\alpha \Gamma^{\alpha \mu \beta} \psi_\beta
= \ol{\psi}^\alpha \theta_{\alpha \mu}(a^*)
\gamma^{\mu \rho \nu} \theta_{\nu \beta}(a) \psi^\beta.
\end{equation}
We can use (\ref{action1}) to write this as
\begin{equation}
\label{oldEM}
\begin{split}
j^\mu &= \ol{\psi}_\beta \gamma^\mu \psi^\beta
- A_1 \left(\ol{\psi} \cdot \gamma \right) \psi^\mu \\
&\quad - A_1 \ol{\psi}^\mu \left(\gamma \cdot \psi \right)
+ A_3 \left(\ol{\psi} \cdot \gamma \right) \gamma^\mu
\left(\gamma \cdot \psi \right),
\end{split}
\end{equation}
where we have used $A_1 = A_2$ since the parameter is now real.
The definitions of $\chi_1$ and $\chi_2$ allow us to write the
charge as
\begin{equation}
\label{oldEMcharge}
\begin{split}
&Q_{\text{EM}} =  \\
&\; \int d^{d-1} x \left[ \psi_\beta^\dagger \psi^\beta
- A_1 \left( \chi_1^\dagger \chi_2
+ \chi_2^\dagger \chi_1 \right)
+ A_3 \chi_1^\dagger \chi_1 \right]
\end{split}
\end{equation}
and we see that our new symmetry charge (\ref{newEMcharge}) involves only the cross 
terms contained in the usual electromagnetic charge (\ref{oldEMcharge}).

The two currents (\ref{newEM}) and (\ref{oldEM}) are separately 
conserved and hence we can form linear combinations of them to get
other vector currents. Although both currents couple to the lower spin 
components of the vector-spinor field, we can modify how much influence 
each of these lower spins has. 
In fact it may be possible that, with judicious choices of couplings, 
one could eliminate the contribution of one or the other of the lower 
spins altogether by eliminating the cross term which contains both 
spin-$\frac{3}{2}$ and spin-$\frac{1}{2}$ components. 
Using this new freedom, it seems clear that the new current
will have an important influence on the inconsistency problems that have
been found in all interactions involving spin-$\frac{3}{2}$ fields
\cite{johnson1961,velo1969,deser2000}.
Perhaps the inconsistencies can be made to cancel between the
the different conserved currents so that the new symmetry 
may lead to further progress in that long-standing problem. 

\section{Conclusion}

We have presented the most general lagrangian and propagator for the 
Rarita-Schwinger field in $d$ dimensions. These are given by equations 
(\ref{action1}) and (\ref{propagator1}) should prove 
useful in calculating higher loop effects in dimensional regularization
as would occur in the effective resonance contribution to the imaginary 
part of pion scattering amplitudes, anomalous magnetic moments, and 
many other processes for which the $\Delta(1232)$ resonance or any other 
spin-$\frac{3}{2}$ particles play a significant role.

We have also explored the invariance properties of the general action under
rotations of the lower spin, off-shell fields and found that this invariance
implies the existence of a new conserved vector current and charge.
A remarkable and important aspect of our technique is that we have 
used a symmetry coming from a {\it non-unitary} continuous group. 
The new current, in combination with the usual electromagnetic vector current
leads to new (possibly even fully consistent) interactions involving 
spin-$\frac{3}{2}$ fields such as the electromagnetic couplings that we 
have given in (\ref{newEM}) above, couplings to any other vector fields 
such as vector mesons, derivative couplings to scalar fields such as the 
pion, etc. 

\section{Acknowledgments}
I would like to thank Emil Akhmedov, Valery Dolotin, Osvaldo Santillan 
and Vladimir Pascalutsa for helpful comments.


\begin{thebibliography}{50}
%
\bibitem{wetterich1983} C. Wetterich, 
{\it Massless spinors in more than four dimensions}, Nucl. Phys. {\bf B211}, 177
(1983).
%
\bibitem{van1981} Peter van Nieuwenhuizen, 
{\it Supergravity}, Phys. Rep. {\bf 68}, 189 (1981).
%
\bibitem{rarita1941} William Rarita and Julian Schwinger, 
{\it On a theory of particles with half-integral spin},
Phys. Rev. {\bf 60}, 61 (1941).
%
\bibitem{aurilia1969} A. Aurilia and H.  Umezawa, 
{\it Theory of High Spin Fields}, Phys. Rev. {\bf 182}, 1628 (1969).
%
\bibitem{pascalutsa1999} Vladmir Pascalutsa and Rob Timmermans, 
{\it Field theory of nucleon to higher-spin baryon transitions},
Phys. Rev. C {\bf 60}, 042201 (1999).
%
\bibitem{johnson1961} K. Johnson and E.C.G. Sudarshan, 
{\it Inconsistency of the local field theory of charged spin 3/2
particles}, 
Ann. Phys. {\bf 13}, 126 (1961).
%
\bibitem{velo1969} G. Velo and D. Zwanziger, 
{\it Propagation and quantization of Rarita-Schwinger waves in an
external electromagnetic potential}, 
Phys. Rev. {\bf 186}, 1337 (1969).
%
\bibitem{deser2000} S. Deser, A. Waldron and V. Pascalutsa,
{\it Massive spin-$\frac{3}{2}$ electrodynamics}, 
Phys. Rev. D {\bf 62}, 105031 (2000).
%
\bibitem{pilling2004} Terry Pilling, hep-th/0404131.
%
\end{thebibliography}
\end{document}